\documentclass[10pt,a4paper,twocolumn]{article}
\usepackage{titlesec}
\usepackage{color}
\usepackage[utf8]{inputenc} 
\usepackage[affil-it]{authblk}

\usepackage{lipsum}
\newcommand\blfootnote[1]{%
  \begingroup
  \renewcommand\thefootnote{}\footnote{#1}%
  \addtocounter{footnote}{-1}%
  \endgroup
  }
\usepackage{siunitx}
\usepackage{float}
\usepackage{tikz}
\usepackage{xcolor}
\usepackage{comment}

\usepackage{braket}
\usepackage{nicefrac}
\usepackage[english]{babel} 
\usepackage[top=2.8cm,bottom=2.8cm,left=2.cm,right=2.cm]{geometry}  
\usepackage{amsmath,amsfonts,amssymb}  

\usepackage{caption}
\usepackage{subcaption}
\usepackage{graphicx}  
\graphicspath{{./Figures/}{./PSTricks/}}  
\usepackage{multirow}
\usepackage{appendix}
\usepackage{wrapfig}
\usepackage{setspace}

\usepackage{array,booktabs}

\usepackage{enumitem}
\usepackage{caption3} 
\DeclareCaptionOption{parskip}[]{}
\usepackage[font=small,labelfont=bf]{caption}
\usepackage[final]{pdfpages}
\usepackage[colorinlistoftodos]{todonotes}
\usepackage{tikz}
\usepackage{sidecap}
\sidecaptionvpos{figure}{t}
\usepackage{abstract}

\usepackage[pdftex]{hyperref} 

\title{\textbf{Experimental demonstration of the DPTS QKD protocol over a 170 km fiber link.}} 
\author{Beatrice Da Lio*, Davide Bacco$^\dagger$, Daniele Cozzolino, Yunhong Ding, Kjeld Dalgaard,\\ Karsten Rottwitt, Leif Katsuo Oxenløwe}
\affil{\small CoE SPOC, DTU Fotonik, Dep. Photonics Eng., Technical University of Denmark, Orsteds Plads 340, Kgs.~Lyngby, 2800 Denmark}
\date{\vspace{-1em} \small   Dated: \today } 

\begin{document}

\pagestyle{plain}
\setcounter{page}{1}
\twocolumn[ 
\begin{@twocolumnfalse}
\maketitle
     \vspace{-0.8cm}
  \begin{abstract}
      \normalsize
         \vspace*{-1.0em}
\noindent Quantum key distribution (QKD) is a promising technology aiming at solving the security problem arising from the advent of quantum computers. While the main theoretical aspects are well developed today, limited performances, in terms of achievable link distance and secret key rate, are preventing the deployment of this technology on a large scale. More recent QKD protocols, which use multiple degrees of freedom for the encoding of the quantum states, allow an enhancement of the system performances. Here, we present the experimental demonstration of the differential phase-time shifting protocol (DPTS) up to $170$ km of fiber link. We compare its performance with the well-known coherent one-way (COW) and the differential phase shifting (DPS) protocols, demonstrating a higher secret key rate up to $100$ km. Moreover, we propagate a classical signal in the same fiber, proving the compatibility of quantum and classical light.
\end{abstract}
  \end{@twocolumnfalse}
 ]

The\blfootnote{* bdali@fotonik.dtu.dk, $^\dagger$dabac@fotonik.dtu.dk} security of digital data is extremely important in our society, due to the continuous exchange of sensitive information. Classical cryptography is based on mathematical assumptions which do not guarantee information-theoretic security~\cite{Shor1997}, {\it i.e.} a security that cannot be broken with unlimited computational power. However, quantum key distribution (QKD), a branch of quantum communication (QC), provides unconditional security based on the laws of quantum physics~\cite{BBPr,Scarani2009}. 
In the last 30 years, free-space, underwater and fiber based experiments have demonstrated the exploitation of different physical degrees of freedom for QC protocols~\cite{Boaron2018,Ji2017,Bouchard2018,Vallone2015}.
Among these, the differential phase reference (DPR) schemes were proposed as a step towards easier implementations for fiber transmission schemes. They make use of either the time of arrival of pulses, the phase difference between them or, more recently, both dimensions to encode secure key bits~\cite{Bacco2016,DaLio2017,Stucki2005,Inoue2002}.
Furthermore, several quantum networks have already been implemented~\cite{Sasaki2011,Chen2009,Peev2009,Liao2017}. During the last decades, the efforts of the scientific community were focused on enhancing quantum communication performance in terms of key rate, transmission distance and security aspects~\cite{Ma2005,Hwang2003,Bacco2016,Zhong2015,Lo2o12,Yin2016,Islam2017,Lucamarini2018,Roberts2017,Dynes2016}.
Here, we present a practical implementation of the differential phase-time shifting (DPTS) protocol over 170 km of single mode fiber, proving a higher secure key rate compared with other protocols of the DPR family, such as the coherent one-way (COW) and the differential phase shifting (DPS)~\cite{Bacco2016,Stucki2005,Inoue2002}.
Furthermore, we also prove that a classical signal at a different wavelength can coexist on the same optical fiber up to 90 km distance.

\begin{figure*}[t]
\centering
\includegraphics[width=17cm]{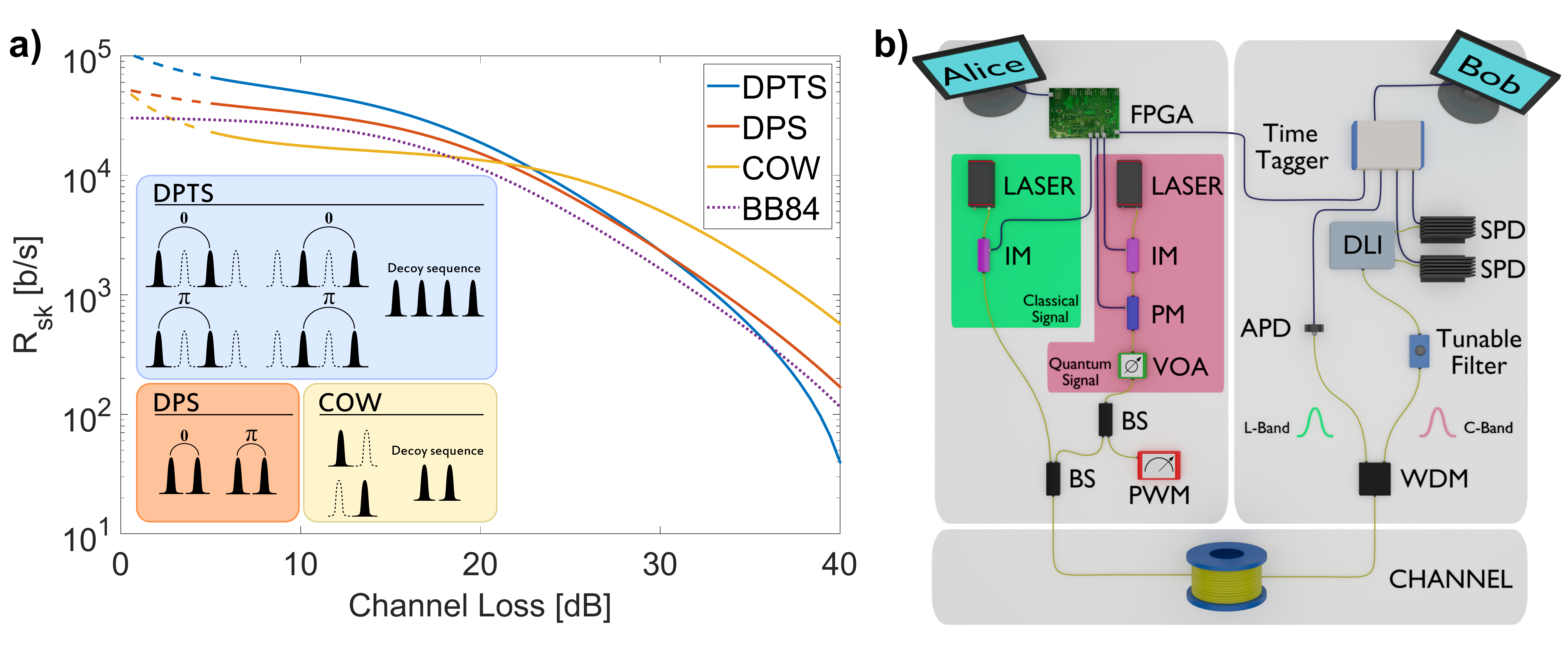}
\caption{{\bf a) Theoretical secret key rates R$_{\textbf{sk}}$ as a function of channel loss.} DPTS (blue), DPS (orange) and COW (yellow) secret key rates under the condition of beam splitting attack and BB84 with decoy state method (dotted violet) secret key rate against collective attacks. Parameters: $\nu=1.19$ GHz, $r_{dc}=100$ Hz, $\eta=20\%$, $t_d=20 \enskip \mu$s, $\mu_{DPTS}=0.26$, $\mu_{DPS}=0.13$, $\mu_{COW}=0.52$, $\mu_{BB84}=0.25$ (signal), $\nu_{BB84}=0.08$ and $\omega_{BB84}=10^{-10}$ (decoys) photon/pulse; $V=0.98$ and $l_{int}=8$ dB (DPTS, DPS and BB84); probability of decoy sequence $p_{d}=0.1$ (DPTS and COW); $N=6$ pulses/block (DPTS). Inset shows the encoding symbols in DPR protocols: filled pulses are WCPs, dotted pulses are vacuum states. {\bf b) Schematics of the experimental setup.} FPGA: field programmable gate array board; continuous wave lasers: 1550 nm for quantum signal (C-band) and 1610 nm for classical signal (L-band); IM: intensity modulator; PM: phase modulator; VOA: variable optical attenuator; BS: beam splitter; PWM: power meter; WDM: wavelength division multiplexer filter; APD: avalanche photodiode; DLI: delay line interferometer; SPD: single photon detector.
}
\label{fig:symbol_setup}
\end{figure*}

The DPTS protocol encodes the information in relative properties of consecutive weak coherent pulses (WCPs).
However, as opposed to the other DPR protocols, the DPTS exploits more than one degree of freedom at once, namely the position in time and the phase difference among consecutive pulses. This allows the DPTS to improve the secret key rate in an intra-city network scenario (in terms of reachable distances and channel loss), while at the same time being more robust against channel noise as shown in Figure~\ref{fig:symbol_setup}a)~\cite{Bacco2016}.
In the DPTS protocol the information is encoded in four possible symbols in the alphabet $\left\lbrace 0,1,2,3 \right\rbrace$, which are:
\begin{eqnarray}
\label{eq:symbols}
&&\ket{0}=\ket{\pm\alpha}\ket{vac}\ket{\pm\alpha}\ket{vac},\nonumber \\
&&\ket{1}=\ket{\pm\alpha}\ket{vac}\ket{\mp\alpha}\ket{vac},\nonumber \\
&&\ket{2}=\ket{vac}\ket{\pm\alpha}\ket{vac}\ket{\pm\alpha},\nonumber \\
&&\ket{3}=\ket{vac}\ket{\pm\alpha}\ket{vac}\ket{\mp\alpha}.
\end{eqnarray}
The terms $\ket{\pm\alpha}$ and $\ket{vac}$ in eq.~\ref{eq:symbols} represent a coherent state of intensity $\alpha$ and a vacuum state respectively. The $\pm$ sign represents the phase of the state. When the two coherent states have the same (opposite) sign, their phase difference is $0$ ($\pi$), see figure~\ref{fig:symbol_setup} a).
When the transmitter, usually called Alice, has prepared the quantum states, she sends them through a quantum channel towards the receiver, called Bob. To measure them, Bob uses a delay line interferometer, with delay $T=2/\nu$ ($\nu$ is the repetition rate), to sort among the $0$ or $\pi$ phase difference. At the same time Bob measures on each output of the interferometer the time of arrival of the pulses.
After the quantum communication process, the sifting procedure takes place.
Hence, Alice and Bob share a sifted key and the following steps in the protocol are given by the classical error correction and privacy amplification.
The equations for the final achievable secret key rate, under the assumption of beam splitting attacks, are reported in the supplementary materials~\cite{Scarani2009,Bacco2016}.
Figure~\ref{fig:symbol_setup}a) shows a comparison of the achievable secret key rate using DPTS, DPS and COW from the DPR family and the secret key generated with the standard BB84 protocol with decoy state method. To be noted that the BB84 protocol offers unconditional security~\cite{Lo2005}, \textit{i.e.} it is secure against collective attacks (the most comprehensive kind of attacks), while the DPR protocols are secure against collective beam splitting attacks. Even though a fair comparison is not possible, it is also important to highlight that the experimental implementation of the BB84 with decoy states requires more equipment, a more advanced control unit to generate all the states in the mutually unbiased bases and also a more complicated receiver.


\begin{figure*}[t]
   \centering
    \includegraphics[width=17cm]{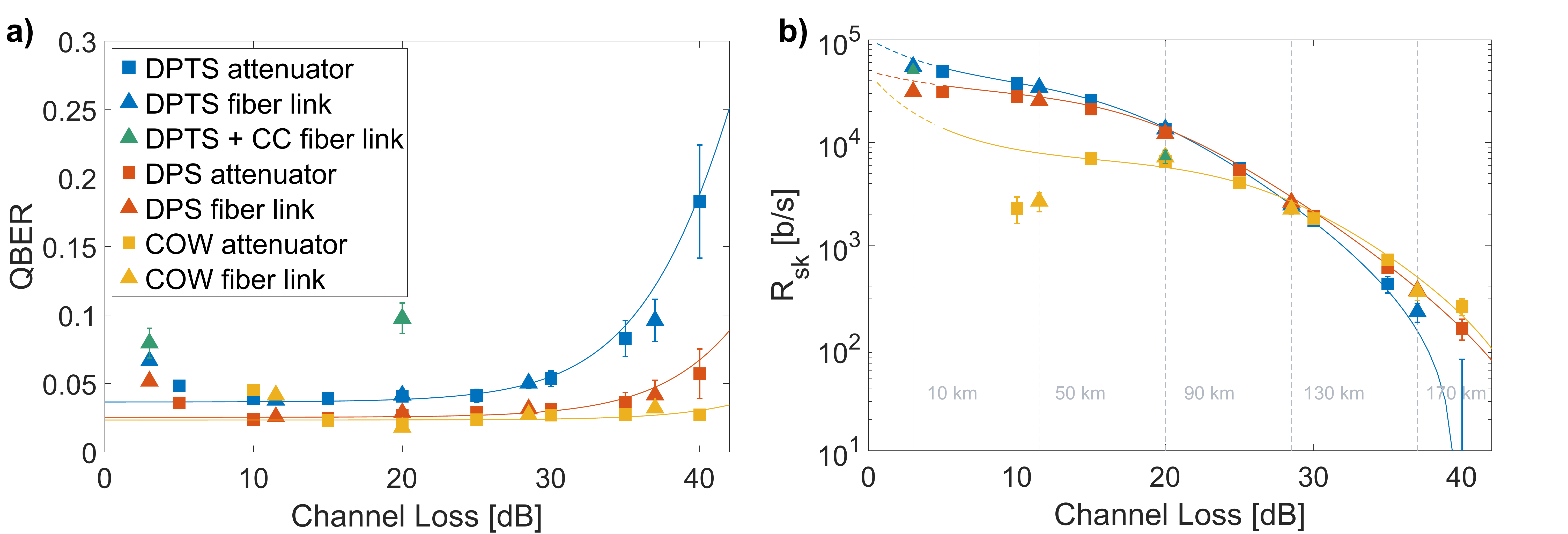}
    \caption{{\bf a) Measured QBER and b) experimental secret key rate.} The color scheme used is: DPTS without co-propagating classical channel (blue), with optical classical synchronization (green), DPS (orange) and COW (yellow). Triangles report data collected using fiber spools as channel link, squares show data obtained emulating channel loss with a VOA. Solid lines show the simulated results taking into account setup imperfections as intrinsic errors.}
\label{fig:results}
\end{figure*}

\begin{figure}[t]
\vspace{0cm}
   \centering
    \includegraphics[width=8.5cm]{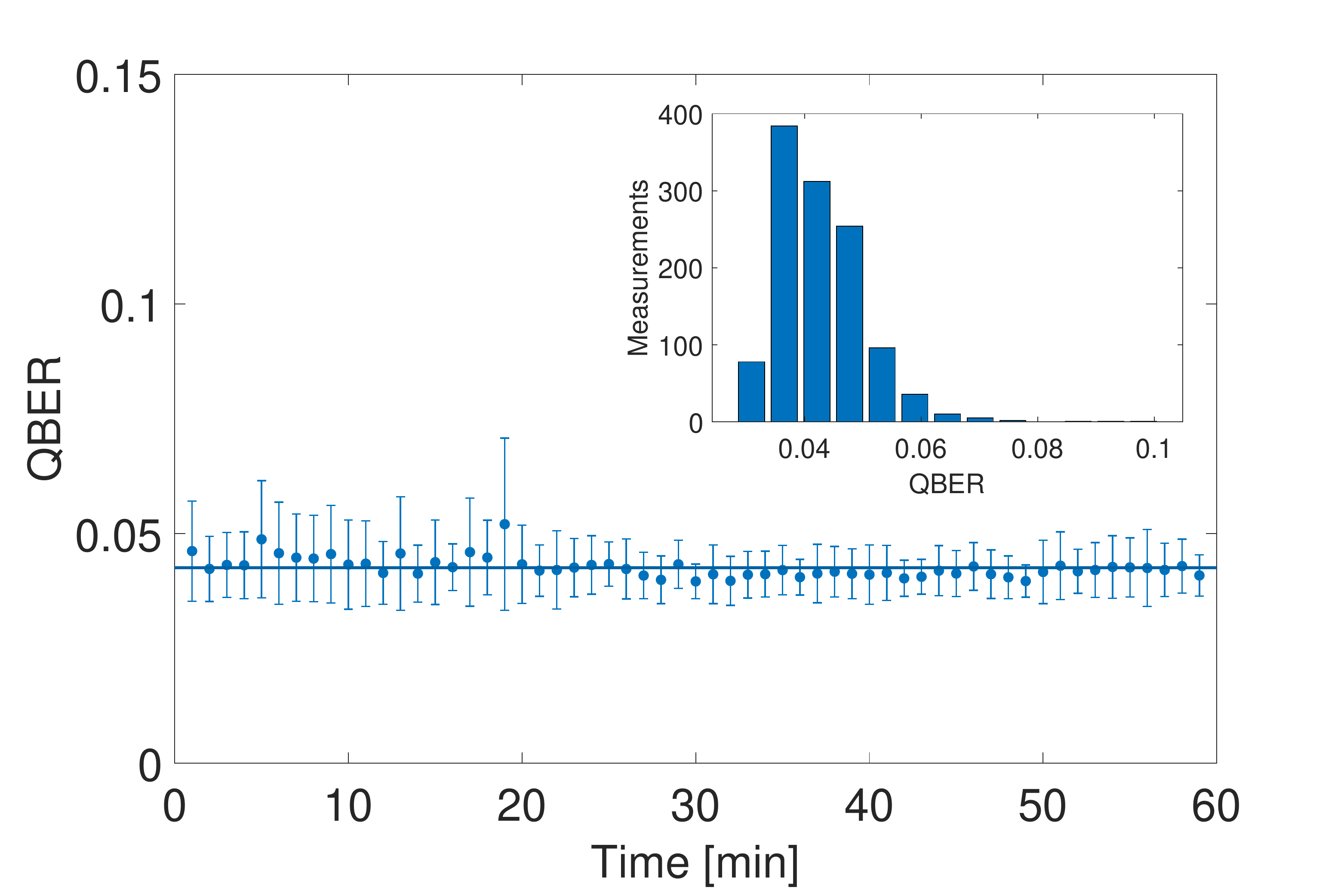}
    \caption{{\bf Stability of the DPTS protocol.} Experimental QBER of the DPTS protocol at 50 km distance for over 1 hour of continuous measurements. Inset shows the QBER distribution.}
    \label{fig:stability}
\end{figure}

The experimental setup used in the current experiment is shown in figure~\ref{fig:symbol_setup}b). To prepare the train of time-encoded WCPs, Alice carves with an intensity modulator a continuous wave laser at $1550$ nm. The obtained signal has an average block length of $N=6$ pulse/block (in a block there are only symbols with the same time encoding~\cite{Bacco2016}). The optical signal is then sent through a phase modulator, which imprints the required phase difference among consecutive WCPs. Hence, Alice uses an optical variable attenuator to reach the quantum regime of $\left|\alpha \right|^2=\mu \approx 0.26$ photon/pulse. With this value the secret key generation rate of the DPTS is maximized~\cite{Bacco2016}. The WCPs are then sent to Bob through a single mode fiber link of variable length, from $10$ km to $170$ km. The loss per unit distance of the fiber is $0.22$ dB/km. At the receiver side, Bob uses a free-space delay line interferometer with overall insertion loss of approximately $l_{int}\approx 8$ dB and visibility $V\approx 0.98$ to infer the quantum states. A phase difference of $0$ is routed towards one output of the interferometer, while a phase difference of $\pi$ constructively interferes on the other output. The two outputs are then linked to two IDQ230 InGaAs single photon detectors, which have the following parameters: efficiency $\eta_{det}=20\%$, dark count rate $r_{dc}\approx 100$ Hz, dead time $t_d=20 \quad \mu$s and jitter $t_j \approx 300$ ps. Both detectors are connected to a time tagging unit. 
The electrical control at Alice's side is given by a field programmable gate array (FPGA) board whose three electrical outputs are for the intensity modulator, for the phase modulator and the synchronization signal. The repetition frequency is $\nu = 1.19$ GHz, the electrical pulse width is of approximately $100$ ps whereas the obtained optical pulse width is around $150$ ps. The synchronization signal is either sent electrically to Bob's time-tagger unit directly from the FPGA or converted into a classical optical signal which co-propagates with the quantum channel. More details are reported in the supplementary materials.
To implement the DPS and COW protocols, the setup is easily adapted. We changed the delay of the optical interferometer in order to have a fair comparison between the protocols, {\it i.e.} all the protocols are implemented at the same transmitter speed. Moreover, for the COW protocol, at Bob's side an unbalanced beam splitter is required so that most of the time the pulses are directly received using one single photon detector but sometimes, with a low probability, they are used to check coherence and therefore sent to the delay line interferometer. For both protocols, the optimal mean photon number per pulse is used in the implementation~\cite{Bacco2016,Branciard2008}.

Figure~\ref{fig:results} shows the performance comparison of the three protocols in terms of quantum bit error rate (QBER) and secret key rate under the condition of beam splitting attacks (when the synchronization signal is electrically sent to Bob). The triangles represent data collected for each protocol when the link between Alice and Bob is made by fiber spools (with distances ranging from 10 to 170 km with steps of 40 km), whereas the squares are measurements taken when an optical attenuator constitutes the channel, thus only emulating fiber loss (from 5 to 40 dB losses with steps of 5 dB). Solid curves represent simulations taking setup imperfections into account. An intrinsic error of $e_{t}=1.5\%$ is estimated in the time domain, due to the finite extinction ratio of the intensity modulator during the carving procedure (for COW and DPTS protocols), and an intrinsic error of $e_{p}=0.5\%$ affects the phase domain, due to an imperfect modulation in the phase modulator (for DPTS and DPS protocols). 
Note that the bounds used in this work to compute the achievable secret key rate are valid in the long distance regime~\cite{Branciard2008,Bacco2016}, which is assured after approximately 5 dB of channel loss (corresponding to a $\sim 23$ km link). This is also shown in figure~\ref{fig:results}b), where the simulation curves are dashed before reaching this regime.
Finally, as a preliminary demonstration of our system used in real communication networks, we co-propagate a classical channel, carrying the optical synchronization used for the QKD system, with the DPTS signal. This is obtained by modulating a CW laser at 1610 nm. At the receiver side, a first wavelength division multiplexing filter (extinction ratio $60$ dB) separates the two signals directing the synchronization to a photo detector. The quantum channel is further filtered with a narrow band-pass filter with 3-dB bandwidth of $0.8$ nm around 1550 nm and extinction ratio of 40 dB. The second filtering step is needed to further reduce the leakage from the classical channel into the quantum channel.
The results are reported in figure~\ref{fig:results}a) and b) with green triangle markers. The input power of the classical channel was set to $-27$ dBm, minimizing the impact on the quantum channel but ensuring enough power for a successful detection with the photodiode.

The comparison of the three protocols shows that for short-range links, \textit{i.e.} up to 21 dB channel loss, the achievable secret key rate is indeed higher when using the DPTS compared to DPS and COW performances.
Note that the experimental values up to 5 dB attenuation, for the DPTS and the DPS protocol, exhibit a higher QBER than expected. This is mainly due to the detectors' saturation regime. For the COW protocol, which does not rely on an interferometer with insertion loss at Bob's side, the detectors saturate up to 11 dB (50 km) of channel loss.
An interesting channel distance to consider is thus 50 km, where the bounds conditions are valid. Here, the DPTS reaches a secure rate of 34 kb/s with a system that is stable for over one hour, as reported in figure~\ref{fig:stability}. The DPS protocol is able to produce 25 kb/s of secret key rate and the COW protocol produces only 2.7 kb/s in the experimental implementation, while the simulation curve reaches up to 7.8 kb/s (saturation regime).
The DPTS protocol shows indeed an improved performance in the secret key rate and a better robustness against noise for applications in an intra-city scenario. On the other hand, on longer distances the secret key rate drops more rapidly than the other protocols, even though a positive secret key rate can still be experimentally obtained for a distance of 170 km. An intuitive explanation to this is given by the fact that any error can affect both time and phase domains: when the random dark count clicks are comparable in number to the actual photon clicks, then this effect starts having a more severe impact on the protocol performance.
In the case of co-propagation of classical and quantum signal, the higher QBER and the respective decrease of the secret key rate result from leakage from the classical channel and the detectability of the classical channel itself. Indeed, the information the classical signal is carrying is crucially needed for synchronizing the quantum channel. The maximum distance we could achieve was 90 km with a secret key rate of 7.3 kb/s. To increase the transmission distance it appears necessary to introduce a more sophisticated filtering scheme, which would allow higher classical input power, and/or amplification schemes for the classical channel.

In this paper, we demonstrated the DPTS protocol over 170 km of single mode fiber and compared its performance with other DPR protocols. We showed that in an intra-city network scenario, the DPTS outperforms the other protocols for up to 21 dB channel loss (about 100 km) under the assumption of beam splitting attacks. We also demonstrated that our scheme can co-exist on the same fiber with classical light, necessary for a complete deployment of QKD systems.

\subsection*{Acknowledgements}
This work is supported by the Centre of Excellence, SPOC (Silicon Photonics for Optical Communications (ref DNRF123) and from the People Programme (Marie Curie Actions) of the European Union's Seventh Framework Programme (FP7/2007-2013) under REA grant agreement n$^\circ$ $609405$ (COFUNDPostdocDTU).

\subsection*{Author contributions}
D. Bacco proposed the idea. D. Bacco and B. Da Lio carried out the theoretical analysis on the proposed protocol. B. Da Lio and D. Bacco carried out the experimental work. D. Bacco, B. Da Lio, D. Cozzolino, Y. Ding, K. Rottwitt, and L. K. Oxenløwe discussed the results. All authors contributed to the writing of the manuscript.

\subsection*{Competing financial interests}
The authors declare no competing financial interests.

\nocite{*}
\bibliographystyle{unsrt}
\bibliography{bibliography}

\end{document}